\documentclass[11pt]{article}
\usepackage{mons}
\usepackage{graphicx}
\usepackage{times}

\newcommand{\al}{\rm et al.} 
\newcommand{\kms}{km s$^{-1}$}

\lefthead{Lastennet et al.}
\righthead{Preparation of the COROT space mission}

\begin{document}

\title{Preparation of the COROT space mission: 
       fundamental stellar parameters determinations 
       from photometric and spectroscopic analyses 
       for COROT potential target candidates}

\author{E. Lastennet}
\affil{Depto de Astronomia, UFRJ,
       Ladeira Pedro Ant\^onio 43, 20080-090 Rio de Janeiro RJ, Brazil }

\author{F. Ligni\`eres}
\affil{Laboratoire d'Astrophysique, CNRS UMR 5572,
       OMP, 57, av. d'Azereix, 65008 Tarbes Cedex, France}

\author{R. Buser}
\affil{Astronomisches Institut der Universit\"at Basel, Venusstr. 7,
       CH-4102 Binningen, Switzerland}

\author{Th. Lejeune}
\affil{Observat\'orio Astron\'omico de Coimbra, Santa Clara,
           P-3040 Coimbra, Portugal}

\author{Th. L\"uftinger}
\affil{Institut f\"ur Astronomie, Universit\"at Wien,
       T\"urkenschanzstr. 17, A-1180 Wien, Austria}

\author{F. Cuisinier}
\affil{Depto de Astronomia, UFRJ,
       Ladeira Pedro Ant\^onio 43, 20080-090 Rio de Janeiro RJ, Brazil}

\author{C. van 't Veer-Menneret}
\affil{Observatoire de Paris, 61 avenue de l'Observatoire,
       F-75014 Paris, France}

\begin{abstract}

We present a sample of 9 nearby F-type stars with detailed spectroscopic
analyses to investigate the Basel Stellar Library (BaSeL) in two photometric 
systems simultaneously, UBV Johnson and $uvby$ Str\"omgren.
The sample corresponds to potential targets of the central seismology programme
of the COROT (COnvection \& ROtation) space experiment, which have been recently 
observed at Observatoire de Haute-Provence (OHP, France).  
The atmospheric parameters T$_{\rm eff}$, [Fe/H], and log g obtained from the BaSeL
models are compared with spectroscopic determinations as well as with results of 
other photometric calibrations (the ``Templogg'' method and the catalogue of Marsakov 
\& Shevelev, 1995). 
Moreover, new rotational velocity determinations are also derived from the 
spectroscopic analysis and compared with previous results compiled in the SIMBAD 
database.  
For a careful interpretation of the BaSeL solutions, we computed confidence regions
around the best $\chi$$^2$-estimates and projected them on T$_{\rm eff}$-[Fe/H],
T$_{\rm eff}$-log g, and log g-[Fe/H] diagrams. 
In order to simultaneously and accurately determine the stellar parameters
T$_{\rm eff}$, [Fe/H] and log g, we suggest to use the combination of the synthetic BaSeL
indices B$-$V, U$-$B and b$-$y (rather than the full photometric information 
available for these stars: B$-$V, U$-$B, b$-$y, m$_1$ and c$_1$) and we present full 
results in 3 different diagrams, along with the results of other methods (photometric 
and spectroscopic). All the methods presented give consistent solutions, and 
the agreement between Templogg and BaSeL for the hottest stars of the sample could 
be especially useful in view of the well known difficulty of spectroscopic 
determinations for fast rotating stars. 
Finally, we present current and future developments of the BaSeL models for 
a systematic application to all the COROT targets. 
   
\end{abstract}

\section{Introduction}

In order to prepare the target selection for the asteroseismology space
missions MONS and COROT, a photometric and spectroscopic survey is
planned of bright main target candidates as well as fainter targets to be
observed. 
The data obtained from different sites will be used for an automatic
classification to derive effective temperature (T$_{\rm eff}$), gravity (log g), 
and metallicity ([Fe/H]) of stars. 

During the COROT-MONS Workshop, existing tools which must be applied 
for these tasks, and which should be adapted and further
developed to give the maximal support to the space missions through
groundbased support were presented. In this contribution, we 
only focus on the COROT (COnvection and ROTation) 
mission and the results we obtained from 
different methods: a detailed spectroscopic study, and 3 methods 
based on various photometric calibrations. 
One of this method (the BaSeL library) can be easily automatized for future
applications for the COROT targets stars, so we put particular emphasis 
on this method in the context of the present workshop. 

In this contribution, we first present the sample (Section 2) and the different 
methods (Sect. 3) used to assess the validity of the BaSeL library for  
the present purpose.  
In Sect. 4 we discuss the results derived from the BaSeL models, with 
comparison with the other methods presented in Sect. 3. 
Finally in Sect. 5 we present future development of the BaSeL models for 
the COROT target stars, and Sect. 6 draws the general conclusions. 
  
\section{The sample: 9 F-type stars}

We use a sample of 9 nearby ($\leq$ 45 pc) F-type stars for which a detailed 
spectroscopic analysis has been performed.
These stars are potential candidates of the COROT space experiment
main programme (see for instance Catala {\al} 1995, Mi\-chel {\al} 1998, and 
Baglin {\al} 1998 for details), and has been observed at the 193cm telescope 
at Observatoire de Haute-Provence (OHP, France) as part of the target selection 
process.
High signal to noise ratio (S/N $\simeq$ 150) spectra have been obtained using
{\it Elodie} echelle spectrograph and analysed by comparison with theoretical 
spectra (Ligni\`eres {\al} 1999).
This set of new spectroscopic data together with both Johnson and Str\"omgren 
photometry from the literature provides a unique test of the predicted results 
of the BaSeL library. 
In addition, other calibration methods, namely the Templogg programme (based on 
Napiwotzki {\al} 1993, and K\"unzli {\al} 1997) and the Marsakov \& Shevelev
(1995) catalogue, provide further comparisons. \\        

The photometric data in the Johnson (B$-$V and U$-$B) and Str\"omgren
(b$-$y, m$_1$$=$(v$-$b)$-$(b$-$y), c$_1$$=$(u$-$v)$-$(v$-$b)) systems of our
working sample of F-type stars are presented in Table 1, along with
cross-identifications, and Hipparcos parallaxes ($\pi$). 
The issue of reddening is discussed in detail in Lastennet et al. (2001a).  

\begin{table*}[htb]
\caption[]{Cross-identifications (HD and HIP numbers), Hipparcos parallaxes 
(ESA catalogue, 1997; see also Perryman et al., 1997), 
and photometric data used in the
BaSeL and Templogg  determinations for the 9 target stars (accuracy of
photometric data is given in Lastennet et al., 2001a).}   
\begin{flushleft}
\begin{center} \begin{tabular}{rrrlrrrrr}
\hline
\noalign{\smallskip}
 ID$^{\dag}$ & HD  &  HIP    & $\pi$  & B$-$V & U$-$B & b$-$y & m$_1$  & c$_1$ \\
   &  &    & (mas) &              &   & &  &  \\
\noalign{\smallskip}
\hline \noalign{\smallskip}
1 &  43587  &   29860  & 51.76$\pm$0.78 & 0.61  & 0.10 &  0.384 & 0.187& 0.349  \\
2 &  43318  &   29716  & 28.02$\pm$0.76 & 0.49 & 0.00 &  0.322 & 0.154 & 0.446  \\
3 &  45067  &   30545  & 30.22$\pm$0.92 & 0.56  & 0.07 &  0.361 & 0.168 & 0.396  \\
4 &  49933  &   32851  & 33.45$\pm$0.84 & 0.39  & $-$0.09 &  0.270& 0.127 & 0.460   \\
5 &  49434  &   32617  & 24.95$\pm$0.75 & 0.295 & 0.05 &  0.178 & 0.178 & 0.717 \\
6 &  46304  &   31167  & 23.13$\pm$0.76 & 0.25  & 0.06 &  0.158 & 0.175 & 0.816  \\
7 &  162917 &   87558  & 31.87$\pm$0.77 & 0.42  & $-$0.03 &  0.280 & 0.166 & 0.458  \\
8 &  171834 &   91237  & 31.53$\pm$0.75 & 0.37  & $-$0.04 &  0.254 & 0.145 & 0.560  \\
9 &  164259 &   88175  & 43.11$\pm$0.75 & 0.38  & $-$0.01 &  0.253 & 0.153 & 0.560    \\          
\noalign{\smallskip}\hline
\end{tabular}
\end{center}
$^{\dag}$ Arbitrary running number. \\
\end{flushleft}
\end{table*}

\section{Description of the methods}

In this section, we briefly present the methods we used to derive T$_{\rm eff}$, 
log g and [Fe/H]: the BaSeL library, two photometric calibrations and a detailed 
spectroscopic analysis. 

\subsection{The BaSeL library} 

The Basel Stellar Library (BaSeL models) is constituted of the merging of various
synthetic stellar spectra libraries, with the purpose of giving the most comprehensive
coverage of stellar parameters.  
This is a library of theoretical spectra
corrected to provide synthetic colours consistent with empirical colour-temperature
calibrations at all wavelengths from the near-UV to the far-IR (see Lejeune {\al} 1998 
and references therein for a complete description).
It has been corrected for systematic deviations detected in respect to UBVRIJHKLM 
photometry at solar metallicity, and can then be considered as the 
state-of-the-art knowledge of the broad band content of stellar spectra.
These model spectra cover a large range of fundamental parameters
(2000 $\leq$ T$_{\rm eff}$ $\leq$ 50,000 K, $-$5 $\leq$ [Fe/H] $\leq$ 1
and $-$1.02 $\leq$ log g $\leq$ 5.5)
and their photometric calibrations are regularly updated and extended to an 
even larger set of parameters (see e.g. Westera et al. 1999).
\\
The BaSeL library spectra have been calibrated directly for standard dwarf and
giant sequences at solar abundances and using UBVRIJHKLM broad-band photometry, and
are hence expected to provide excellent results in these photometric bands 
(see for instance Lastennet et al., 2001b for a recent applications to AGB 
stars in near-infrared JHKLM photometry).
Since they are based on synthetic spectra, they can in principle be used in many
other photometric systems taken either individually or simultaneously, and this is
another major advantage - used in this work - of these models.                                    

Thus, because the BaSeL library has only been calibrated in the UBVRIJHKLM colours,
the best parameters derivations come from this system.
Whilst parameters derivations in equivalent bandwidth systems such as Washington seem to be
as good with the BaSeL libraries than with empirical methods (Lejeune 1997),
parameter derivations from the BaSeL library in narrow band systems such as Str\"omgren
photometry should be of poorer quality. However, even in this non optimal case, good 
results were obtained for individual stars of detached eclipsing binaries 
(Lastennet et al. 1999). 

In order to derive simultaneously the effective temperature (T$_{\rm eff}$), the
metallicity ([Fe/H]), and the surface gravity (log g) of each star, we minimize a $\chi^2$
-functional (see Lastennet et al. 2001a for all details). 

\subsubsection{The "Templogg" method}

We have also run the "Templogg" program which is designed to determine
effective temperature and log g from either Str\"omgren or Geneva photometry.
For Str\"omgren photometry it uses a Fortran program written by E. Fresno
which relies  upon the grids of Moon \& Dworetsky (1985) in the T$_{\rm
eff}$-log g parameter  space relevant to this paper, with the improvements by
Napiwotzki {\al} (1993).
For Geneva photometry it uses a Fortran program
written by M. Kunzli  (see North {\al} 1994).
The program chooses among eight different regions in the HR diagram for
selecting the best calibration within the Str\"omgren system and three
different regions for the Geneva system.
The results from this method are gathered in Table 3.

\subsubsection{The catalogue of Marsakov \& Shevelev (1995)}

Marsakov \& Shevelev (1995) (hereafter [MS95]) have computed effective temperatures
and surface gravities using Moon's (1985) method, which is also based on the
interpolation of the grids presented in Moon \& Dworetsky (1985).
According to Moon (1985), the standard deviations of the derived
parameters are T$_{\rm eff}$$=$$\pm$ 100 K and log g $=$ $\pm$0.06.
The metallicities of Marsakov \& Shevelev (1995) are obtained with the
equation of Carlberg {\al} (1985) which relates [Fe/H]  with 
the colour excess $\delta$m$_1$ and the Str\"omgren $\beta$ index. 
All the [MS95] results relevant for our sample are given in Tab. 3.

\subsection{Spectroscopic analysis}

The fundamental stellar parameters have been derived from a detailed analysis
of spectra with high signal to noise ratio (S/N $\simeq$ 150) obtained at
OHP with the {\it Elodie} echelle spectrograph (spectra ranging from 3906 \AA \, to
6811 \AA \, at a resolution of $\lambda/\Delta \lambda$ $=$ 42000).
After reduction, the observed spectra were compared with theoretical ones 
constructed from a combination of Kurucz atmospheric models (ATLAS9 - Mixing 
Length Theory of convection with $l/ H_{\rm P} = 0.5$ and without 
overshooting\footnote{This choice of parameters different than those used by Kurucz 
is fully justified in van 't Veer-Menneret \& M\'egessier (1996)}, 
the VALD-2 atomic database (Kupka {\al} 1999),
and the SYNTH radiative transfer  codes (Piskunov 1992) and BALMER9 (Kurucz
1993)). 
In addition, the Least-Squares Deconvolution  method (Donati {\al}
1997) provided accurate determination of the projected rotational  velocities
listed in Tab. 2. While we confirm the low v sin$i$ values of HD 43587, HD 43318, 
HD 45067, HD 49933 and the high v sin$i$ of HD 46304 provided by the SIMBAD 
database, we find slightly largest values (by $\sim$25 to 40\%) for HD 162917, 
HD 171834 and HD 171834. We also derive a new rotational velocity 
determination for HD 49434: v sin$i$ $=$ 79 {\kms}. \\  
Details about these determinations are given in the two
next subsections and the results are summarized in Tables 2 and 3.

\begin{table*}[htb]
\caption[]{Cross-identifications (HD and HIP numbers), and  
 rotational velocities v sin$i$ derived from the application of the LSD method
 of Donati {\al}, 1997 on the OHP spectra (Lastennet {\al} 2001a) and from 
the SIMBAD database. }   
\begin{flushleft}
\begin{center} \begin{tabular}{rrrcc}
\hline
\noalign{\smallskip}
 ID$^{\dag}$ & HD  &  HIP   &  \multicolumn{2}{c}{v sin$i$ ({\kms})}      \\
             &     &        & Lastennet {\al} (2001a)  &  SIMBAD database  \\ 
\noalign{\smallskip}
\hline \noalign{\smallskip}
1 &  43587  &   29860  &   2       & 3 ($<$6)  \\
2 &  43318  &   29716  &   5       & 3 ($<$6)  \\
3 &  45067  &   30545  &   6       & 5 ($<$10) \\
4 &  49933  &   32851  &   10      & 5 ($<$10) \\
5 &  49434  &   32617  &   79      &   \\
6 &  46304  &   31167  &   200     & 200 \\
7 &  162917 &   87558  &   25      & 20  \\
8 &  171834 &   91237  &   64      & 50  \\
9 &  164259 &   88175  &   76      & 54  \\          
\noalign{\smallskip}\hline
\end{tabular}
\end{center}
$^{\dag}$ Arbitrary running number. \\
\end{flushleft}
\end{table*}

\subsubsection{Determination of T$_{\rm eff}$ from the H$\alpha$ line}

The effective temperature can be determined by taking advantage of the sensitivity of
H$\alpha$ line wings. Detailed studies (e.g. van 't Veer-Menneret \&
M\'egessier 1996), have shown that the H$\alpha$ line is independent of the surface gravity
(for non-supergiant stars) for effective temperatures ranging from 5000 K to $\sim$8500 K,
and depends only slightly on the metallicity. T$_{\rm eff}$ is therefore obtained for each
star of the sample by fitting the observed H$\alpha$ line with synthetic spectra computed
from a grid of solar metallicity atmospheric models separated by 250K.

\subsubsection{Determination of the surface gravity and metallicity}

Within the temperature range that we found, we noticed that Fe I absorption lines
depend only on temperature and metallicity, being practically independent of the surface
gravity, while Fe II lines are sensitive to the temperature, metallicity and gravity.
Consequently, the temperature being known from the H$\alpha$ line, the metallicity
along with the microturbulence velocity are determined first by fitting a set of weak
and strong Fe I lines.
Then, the gravity is obtained by fitting Fe II lines.
The spectral region near 6130 \AA \, proved to be suitable for this analysis.
However, the line broadening induced by rotation tends to mix neighbouring lines and
prevent the analysis of individual Fe lines for high values of v sin$i$: 
HD 49434, HD 46304, HD 171834 and HD 164259 (all these stars have 
v sin$i$$\geq$60 {\kms}).                           

\subsection{Other determinations in the literature}

To be as complete as possible, we looked for other determinations available in the literature
and the SIMBAD database.
One of the most comprehensive sources for our purpose is the fifth Edition of the catalogue of
Cayrel de Strobel {\al} (1997), which includes [Fe/H] determinations and atmospheric parameters
($T_{\rm eff}$, log g) obtained from high-resolution spectroscopic observations and detailed
analyses, most of them carried out with the help of model atmospheres.
However, the stars of our sample are not included in this catalogue. Since the catalogue
comprises the literature (700 bibliographical references) up to December 1995,
we only looked for more recent references.
To the best of our knowledge, the catalogue of metallicities of Zakhozhaj
\& Shaparenko (1996) (hereafter [ZS96]) is the only one which contains useful information
for our purpose.
These metallicities are obtained from photometric UBV data and are available for
two stars of our sample: HD 43587 and HD 164259 (see Table 3).

\begin{table*}[htb]
\caption[]{Comparison of fundamental stellar parameters determinations. The {\it Templogg} method
uses Str\"omgren and Geneva photometric data,  Marsakov \& Shevelev 1995 ({\it [MS95]}) used
Str\"omgren data, and the {\it BaSeL} method uses Johnson and Str\"omgren data (B$-$V, 
U$-$B and b$-$y in this table). 
Results from our spectroscopic analysis ({\it Spectro.}), 
and Zakhozhaj \& Shaparenko 1996 {\it [ZS96]} are also given.
The uncertainties are not reported but can be directly seen on the figures.}
\begin{flushleft}
\begin{center}
\begin{tabular}{|l|l|ccccccccc|}
\hline\noalign{\smallskip}
ID$^{\dag}$    &            & 1     & 2      & 3      & 4      & 5      & 6      & 7      & 8      &    9  \\
HD             &            & 43587 & 43318  & 45067  & 49933  & 49434  & 46304  & 162917 & 171834 & 164259 \\
\noalign{\smallskip}\hline \noalign{\smallskip}
               & Method     &  &  &  &  &  &  &  &  &  \\
\noalign{\smallskip}\hline \noalign{\smallskip}
T$_{\rm eff}$  & Templogg    & 6009 & 6420 & 5982 & 6535 & 7321 & 7379 & 6587 & 6714 & 6789 \\
               & [MS95]      & 5952 & 6280 & 6066 & 6625 &      &      & 6629 & 6739 & 6730 \\
               & BaSeL       & 5720 & 6320 & 5940 &
               6600 & 7240 & 7240 & 6660 & 6700 & 6820 \\
               & Spectro.$^{\ddag}$  & 6000 & 6250 & 6000 & 6500 & 7250 &
7250 & 6500 & 6750 & 6750 \\  \noalign{\smallskip}\hline \noalign{\smallskip}
log g          & Templogg    & 4.32 & 4.20 & 4.16 & 4.25 & 4.16 & 3.93 & 4.32 & 4.02 & 4.11 \\
               & [MS95]      & 4.11 & 4.05 & 4.02 & 4.46 &      &      & 4.49 & 4.10 & 4.10 \\
               & BaSeL       & 4.3  & 4.5 & 3.8 & 4.3 & 4.0 &
               3.4 & 4.5 & 3.9 & 4.2 \\
               & Spectro.$^{\ddag}$ & 4.5 & 4.0 & 4.0 & 4.0 & & & 4.0 &  & \\
\noalign{\smallskip}\hline \noalign{\smallskip}
[Fe/H]         & Templogg    & $-$0.13 & $-$0.18 & $-$0.15 & $-$0.48 &
             $-$0.03 & $-$0.09 & $+$0.03 & $-$0.20 & $-$0.11 \\
               & [MS95]      & $-$0.15 & $-$0.18 & $-$0.17 & $-$0.35 &   &   & $+$0.08 & $-$0.15 & $-$0.05 \\
               & BaSeL       & $-$0.2 & $+$0.0 & $-$0.1 & $-$0.6 & $-$0.1 &
               $-$0.8 & $+$0.0 & $-$0.5 & $+$0.0 \\
               & Spectro.$^{\ddag}$  & $-$0.1 & $-$0.3 & $-$0.1 & $-$0.5 & &
& $-$0.2 &  &  \\                 & [ZS96]      & $+$0.03 &   &   &   &   &
&   &   & $-$0.03 \\ \noalign{\smallskip}\hline
\end{tabular}
\end{center}
\small
$^{\dag}$ Running number as in Tab. 1 and 2. \\
$^{\ddag}$ Estimated error: $\Delta$T$_{\rm eff}$$\simeq$$\pm$250 K,
$\Delta$log g$\simeq$$\pm$0.5, $\Delta$[Fe/H]$\simeq$$\pm$0.2. \\ 
\end{flushleft}
\end{table*}
\vspace{-0.5cm}
\normalsize

\section{Results}

\begin{figure*}
\begin{center}
\includegraphics[
width=14cm, height=17.5cm, angle=-90.]{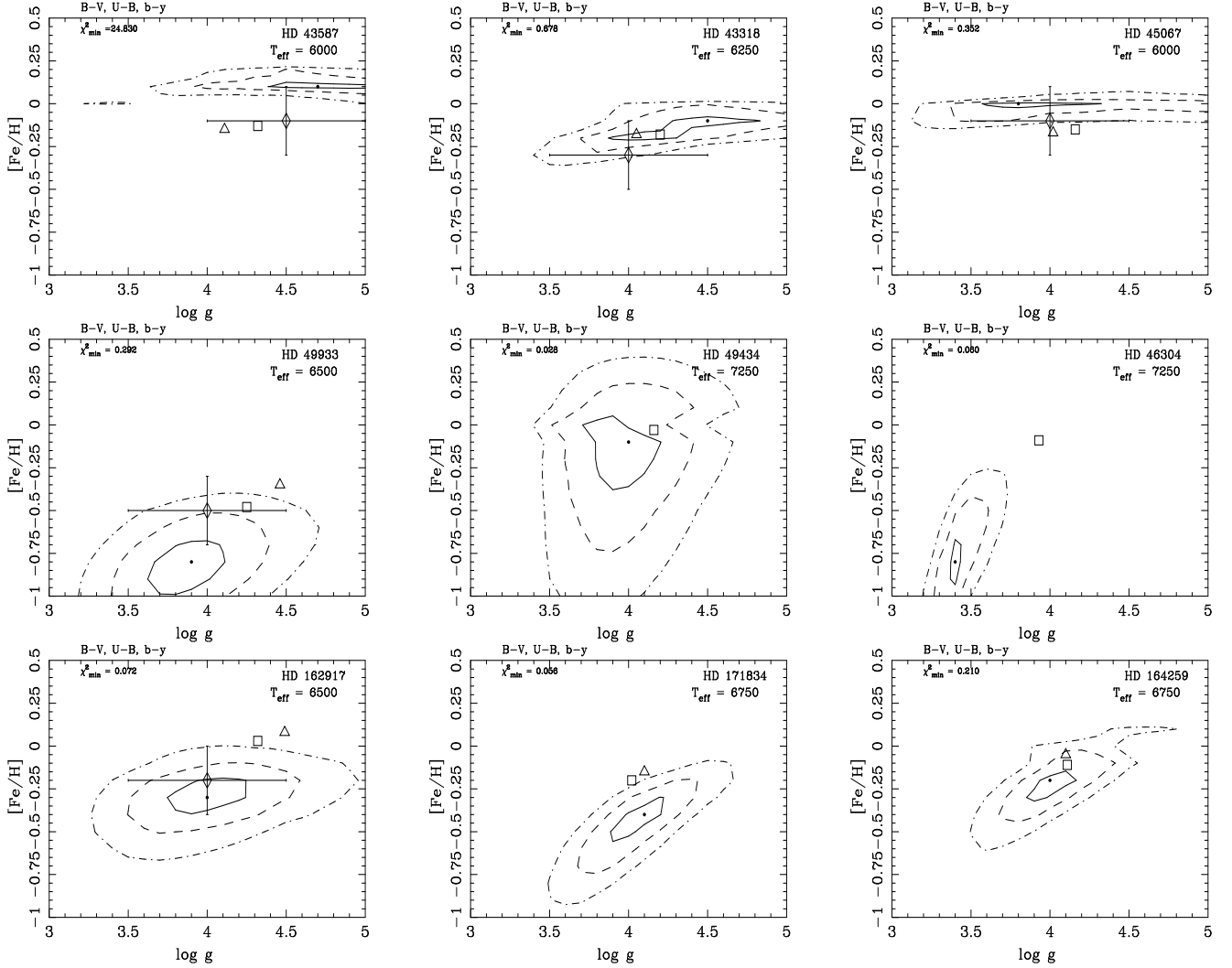}
\end{center}
\caption[]{\label{f:fig1} Simultaneous (log g, [Fe/H]) results for the 9
potential targets of the COROT central seismology programme. The solutions
from the BaSeL models ({\it 1-, 2- and 3-$\sigma$ confidence level contours})
are obtained in order to fit simultaneously the 3 following observed
photometric values : (B$-$V),  (U$-$B) and (b$-$y). 
For each star, the contour solutions
are displayed in a T$_{\rm eff}$ $=$ constant plane, corresponding to the best
T$_{\rm eff}$ spectroscopic estimates. The $\chi^2$$-$value is an estimation
of the quality of the fit (a $\chi^2$-value close to 0 is a good fit). Only
one bad fit is obtained: $\chi^2$$\simeq$25 for HD 43587 ({\it upper left
panel}), but even in this case the solutions are not unrealistic (and at least in 
agreement with the other methods).
 The results projected in the log g-[Fe/H] plane from the
spectroscopic analysis ({\it diamond with error bars}) as well as from the
"Templogg" programme ({\it square}), and Marsakov \& Shevelev (1995) ({\it
triangle}) are also shown for comparison.}
\end{figure*}

\begin{figure*}
\begin{center}
\includegraphics[width=14cm, height=17.5cm, angle=-90.]{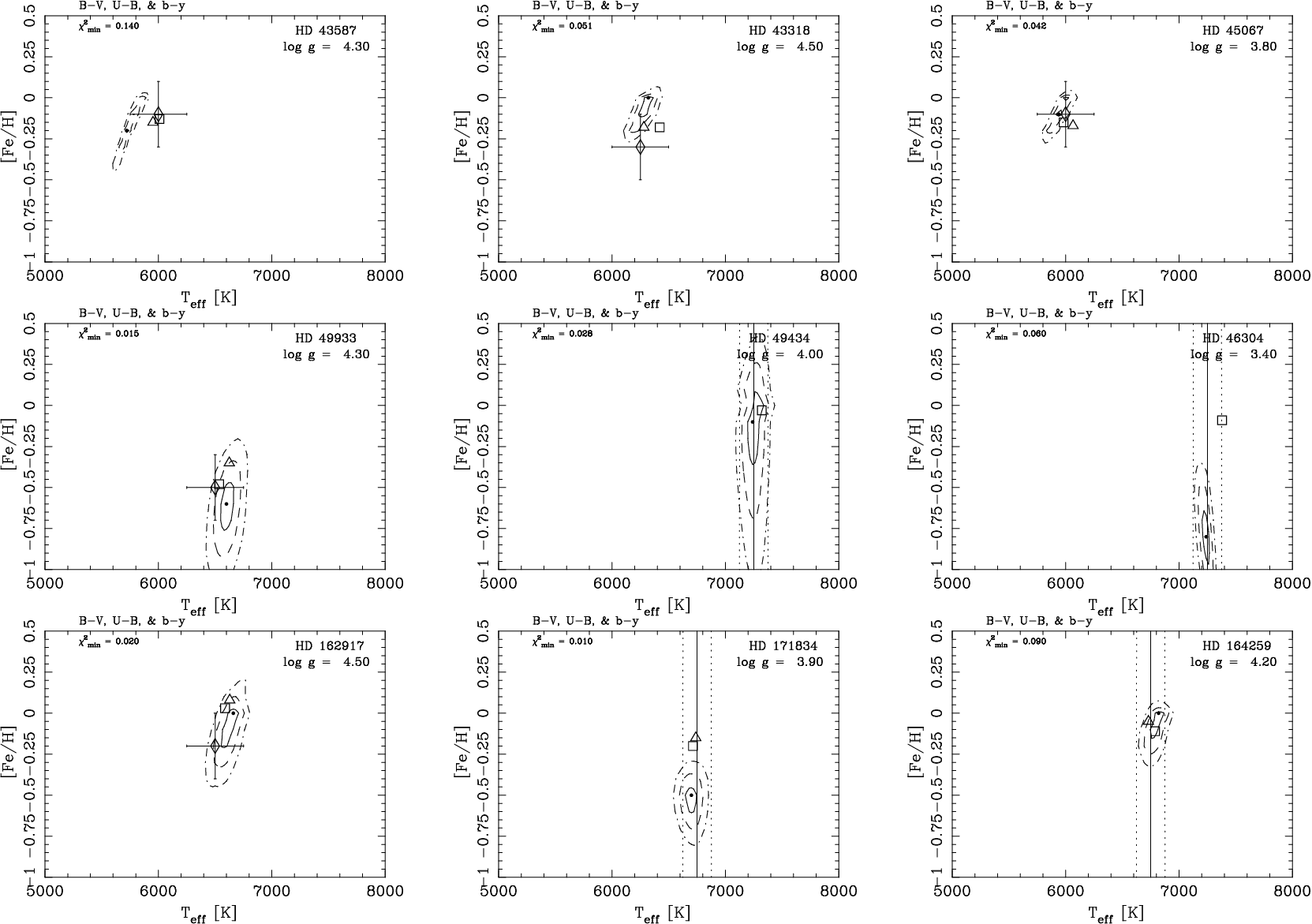}
\end{center}
\caption[]{\label{f:fig2} 
Simultaneous (T$_{\rm eff}$, [Fe/H]) results for the 9 potential targets
of the COROT central seismology programme. 
The solutions from the BaSeL models
({\it 1-, 2- and 3-$\sigma$ confidence level contours}) fitting 
simultaneously the 3 following observed photometric values : (B$-$V),
(U$-$B), (b$-$y). 
For each star, 
the contour solutions are displayed in a log g $=$ constant plane, corresponding
to the best simultaneous (T$_{\rm eff}$, [Fe/H], log g) solutions derived from
the BaSeL models (the grid explored is: 5000 $\leq$ T$_{\rm eff}$ $\leq$ 8000 K
in 20K steps, $-$1 $\leq$ [Fe/H] $\leq$ 0.5 in 0.1 steps, and 3 $\leq$ log g $\leq$ 5
in 0.1 steps).
An estimation of the quality of the best fit ($\chi^2$-value) is also quoted in each panel.
The results projected in the T$_{\rm eff}$-[Fe/H] planes from the
spectroscopic analysis ({\it diamond with error bars}, or {\it solid plus two dotted lines} 
for the hottest stars of the sample)
as well as from the "Templogg" programme ({\it square}),
and Marsakov \& Shevelev (1995) ({\it triangle}) are also shown for
comparison.  
}
\end{figure*}

\begin{figure*}
\begin{center}
\includegraphics[width=14cm, height=17.5cm, angle=-90.]{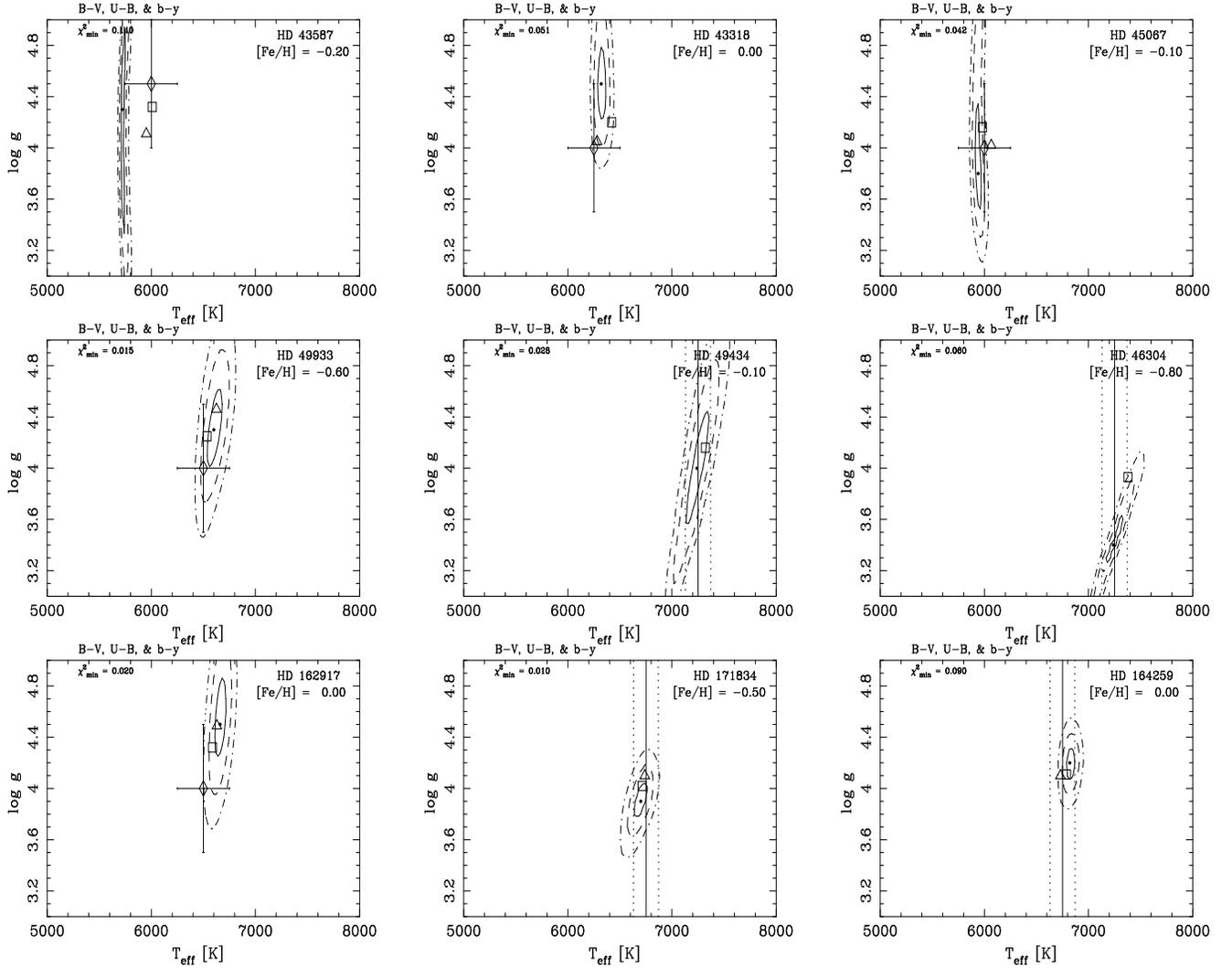}
\end{center}
\caption[]{\label{f:fig3} 
As Fig.~\ref{f:fig2}, but this time in a 
(T$_{\rm eff}$, log g) diagram.
For each star, the contour solutions are displayed in a [Fe/H] $=$ constant plane,
corresponding to the best simultaneous (T$_{\rm eff}$, [Fe/H], log g) solutions 
derived from the BaSeL models. Note that for the hottest stars of the sample 
(HD 46304, HD 49434, HD 164259, and HD 171834), the determination of log g is 
hampered by the line broadening induced by stellar rotation (see Sect. 3.4.2). 
For these stars, the T$_{\rm eff}$ determination is displayed by a {\it solid 
line} plus two {\it dotted lines} because there is no log g determination 
inferred from the spectroscopic analysis. 
}
\end{figure*}

Fig. \ref{f:fig1}, \ref{f:fig2} and \ref{f:fig3} 
show the T$_{\rm eff}$-[Fe/H]-log g results by using the following synthetic 
colour combination: B$-$V, U$-$B, and b$-$y. 
The results are also summarized in Table 3. 
Other photometric combinations are presented and discussed in detailed elsewhere 
(Lastennet et al. 2001a).\\ 
A general comparison with the other methods shows that the
BaSeL solutions are very satisfactory for the 3 fundamental parameters:
effective temperatures are in excellent agreement, and [Fe/H] and gravities show 
good agreement.
The improvement using the B$-$V, U$-$B, b$-$y combination 
is clear when compared with the original results using the B$-$V, U$-$B, b$-$y, 
m$_1$, c$_1$ combination presented in Figs.
1 and 2 of Lastennet et al . (2001a) because there is no longer any systematic
trend towards lower temperatures and because the determinations
of log g and [Fe/H] are  also much better, without systematic deviations. \\
There is only one exception (HD 46304): its T$_{\rm eff}$ is in perfect agreement with
its spectroscopic value, but the predicted log g and [Fe/H] are still low.
While the BaSeL contours are consistent with the Templogg gravity only at the 2-$\sigma$ level,
the metallicity predicted from the BaSeL models is poor in comparison to the result of the Templogg
calibration ($\Delta$[Fe/H]$\simeq$0.7).
What could explain this persistent difference ?
It is worth noticing that this star has a large v sin$i$ (200 km s$^{-1}$, the
largest in our  working sample), and it is well known that high rotational
velocities modify the colours. Since the expected colour effect due to             
rotation is typically a few hundredths of  a magnitude\footnote{These values
are highly dependent of spectral type, age, and chemical  composition, see for
instance Maeder (1971) and Zorec (1992).}  in B$-$V and increases with v sin$i$,
this is probably part of the reason why the predictions of the  BaSeL models
disagree with the Templogg method for this star. However this explanation 
is not satisfying because  rotation is not
taken into account in the Templogg method.  \\ 
In conclusion, except in the
case discussed before, reliable and simultaneous estimates of the 3
atmospheric parameters can be derived for F-type stars from only the three
synthetic BaSeL colours,  B$-$V, U$-$B and b$-$y.  This is a very useful
criterion for further applications.

A comment on the impressively good accuracy of the BaSeL
determinations for the Johnson-Str\"om\-gren B$-$V, U$-$B, b$-$y combination
(see Figs. \ref{f:fig1}, \ref{f:fig2} and \ref{f:fig3}) is in order. \\
The Str\"omgren b$-$y index provides a reliable measure of the
continuum and, therefore, a good temperature index, as discussed in detail in 
Lastennet et al. (2001a). 
This is particularly interesting, given the known difficulties of matching empirical
and theoretical B$-$V-T$_{\rm eff}$ scales to within better than about 0.03 mag
(see, e.g., Sekiguchi \& Fukugita, 2000)\footnote{
The b$-$y index is less vulnerable than B$-$V to the secondary effects of surface
gravity and metallicity, even at the low spectral resolution given by the model spectra.
}.
It is also well known that if one does not ask UBV data to provide the temperature in the first
place (e.g., by using an independent source, such as spectroscopy, spectral classification,
or else suitable other photometry, such as Str\"omgren b$-$y), the sensitivities of
both U$-$B and B$-$V can be used to full advantage for determining log g and [Fe/H].
Moreover, although these sensitivities change with temperature, they
are near or even at their maxima in the F-dwarf star domain (e.g., Buser \& Kurucz 1992).
This means that the derived values of [Fe/H] and log g presented in
Figs. \ref{f:fig1}, \ref{f:fig2} and \ref{f:fig3} are as reliable as they can
possibly be, given the uncertainties in the colours.

\section{Future development of the BaSel models: application for COROT target
stars}

Usually, the BaSeL models provide colours in selected photometric systems,  
given a set of effective temperature, surface gravity and metallicity 
(T$_{\rm eff}$, log g, [Fe/H]). 
The new "BaSeL interactive server", the web version of the BaSeL models
hosted by the Coimbra Observatory since the end of 2000 ({\tt
http://tangerine.astro.mat.uc.pt/BaSeL/}) already provides such 
information.  
We propose to develop in the near future a basic interactive tool based 
on the inverse method: an automatic tool
based on the method of Lastennet et al. (1999) to complete the facilities of
the "BaSeL interactive server". 
This method would be applied to the $\sim$ 1000 remaining potential targets 
of the COROT exploratory programme to provide their fundamental stellar parameters 
as discussed by Lastennet, Lejeune \& Cuisinier (2001).

\section{Conclusion}
Several methods of determination of the fundamental stellar parameters T$_{\rm eff}$,
log g and [Fe/H] are compared for nine single F stars COROT potential targets.
Particular attention has been paid to the simultaneous predictions of the BaSeL models
in two photometric systems, Johnson and Str\"omgren.

We presented the best combination to determine stellar parameters within the BaSeL
library, using a combination of two Johnson colours,     
U$-$B and B$-$V, and a Str\"omgren colour, b$-$y. \\
This BaSeL combination is the best because on the one hand the b$-$y synthetic index
gives reliable and accurate estimates of the effective temperature and, on the other hand,
B$-$V and U$-$B give good estimates of [Fe/H] and the surface gravity.

We also note that the agreement between Templogg and BaSeL for the hottest stars of the
sample could be especially useful in view of the well known difficulty of spectroscopic
determinations for fast rotating stars. \\
As far as the astrophysical applications are concerned, the BaSeL synthetic colours are
of particular interest in evolutionary synthesis and colour-magnitude diagram studies
of stellar populations, such as open clusters and young associations, where
(i) F-type stars are highly common and, of course,
(ii) a vast abundance of data are available in B$-$V, U$-$B and b$-$y colours.
Concerning the determination of the atmospheric parameters of the COROT potential
targets, a result of the present analysis is that all the methods presented 
give consistent solutions.
In the context of the further $\sim$1000 potential targets of the COROT exploratory programme,
it will be of first interest to compare the results of the BaSeL models with those of
other automated spectral analysis methods (e.g. Katz {\al} 1998, and Bailer-Jones 2000).     

\subsection*{Acknowledgments}

We thank J.C. Bouret, C. Catala and D. Katz for their participation in the 
spectral caracterization of the stars. 
This research has made use of the SIMBAD database operated at CDS, Strasbourg 
(France).


\begin{thebibliography}{}

\bibitem[]{1} Baglin A., {\al}, 1998, IAU Symposium 185,
{\it New Eyes to See Inside the Sun and Stars}, eds. F.-L. Deubner, J.
Christensen-Dalsgaard, and D. Kurtz,
p.301

\bibitem[]{2} Bailer-Jones C.A.L., 2000, \aap, 357, 197 

\bibitem[]{5} Buser R., Kurucz R.L., 1992, \aap, 264, 557

\bibitem[]{6} Carlberg R.G., Dawson P.C., Hsu T., Vandenberg D.A., 1985,
\apj, 294, 674

\bibitem[]{7} Catala C., Mangeney A., Gautier D., Auvergne M., Baglin A.,
Goupil M.J., Michel E., Zahn J.P., Magnan A., Vuillemin A., Boumier, P.,
Gabriel A., Lemaire  P., Turck-Chieze S., Dzitko  H., Mosser B.,
Bonneau F., 1995, Astronomical Society of the Pacific Conf. Series,
Vol. 76, p.426

\bibitem[]{8} Cayrel de Strobel G., Soubiran C., Friel E.D., Ralite N., Francois P., 1997,
\aaps, 124, 299

\bibitem[]{11} Donati J.-F., Semel M., Carter B.D., Rees D.E., Cameron A.C., 1997, 
\mnras, 291, 658

\bibitem[]{12} ESA, 1997, {\it The Hipparcos and Tycho Catalogues} (ESA-SP 1200)

\bibitem[]{13} Katz D., Soubiran C., Cayrel R., Adda M., Cautain R., 1998,
\aap, 338, 151

\bibitem[]{14} K\"unzli M., North P., Kurucz R.L., Nicolet B., 1997,
\aaps, 122, 51

\bibitem[]{15} Kupka F., Piskunov N., Ryabchikova T.A., Stempels H.C., Weiss
W.W., 1999, \aaps, 138, 119

\bibitem[]{17} Kurucz R.L., 1993, CD-ROM 13, 14

\bibitem[]{17a}  Lastennet E., Lejeune T., Cuisinier F., 2001, to appear in
    the ASP Conference Series, {\it Observed HR diagrams and stellar
    evolution: the interplay between observational constraints and
    theory}, Coimbra, Portugal

\bibitem[]{19} Lastennet E., Lejeune T., Westera P., Buser R., 1999,
\aap, 341, 857

\bibitem[]{19a} Lastennet E., Ligni\`eres F., Buser R., Lejeune T., L\"uftinger Th., 
          Cuisinier F., van 't Veer-Menneret C., 
          2001a, \aap, 365, 535

\bibitem[]{19c} Lastennet E., Lorenz-Martins S., Cuisinier F., Lejeune T., 2001b, 
            to appear in the ASP Conf. Series, {\it Observed HR diagrams and stellar
            evolution: the interplay between observational constraints and
            theory}, Coimbra, Portugal 

\bibitem[]{21} Lejeune T., 1997, Ph.D. Thesis, Observatoire Astronomique de Strasbourg, France

\bibitem[]{23} Lejeune T., Cuisinier F., Buser R., 1998, \aaps, 130, 65

\bibitem[]{25} Ligni\`eres F., Catala C., Katz D., Lastennet E., L\"uftinger Th., van 't Veer-Menneret C.,
{\it Joint European and National Astronomical Meeting (JENAM 99)}, 7-11 Sept. 1999, Toulouse
(France)

\bibitem[]{26} Maeder A., 1971, \aap, 10, 354

\bibitem[]{27} Marsakov V.A., Shevelev Y.G., 1995, Bull. Inform. CDS 47,13 [MS95]

\bibitem[]{28} Michel E., Baglin  A., {\al}, 1998, {\it Structure and Dynamics of the
Interior of the Sun and Sun-like Stars}, SOHO 6/GONG 98 Workshop Abstract,
June 1-4, 1998, Boston, Massachusetts

\bibitem[]{29} Moon T.T., 1985, Com. Univ. London Obs. 78, 1

\bibitem[]{30} Moon T.T., Dworetsky M.M., 1985, \mnras, 217, 305

\bibitem[]{31} Napiwotzki R, Sch\"onberner D., Wenske V., 1993, \aap, 268, 653

\bibitem[]{32} North P., K\"unzli M., Nicolet B., 1994, 22nd GA of the IAU, the Hague

\bibitem[]{33} Perryman M.A.C., et al. 1997, \aap, 323, L49,  

\bibitem[]{34} Piskunov N.E., 1992, in {\it Stellar magnetism}, Y.V. Glagolevskij, I.I. Romanyuk (eds.),
Nauka, St. Petersburg, p.92

\bibitem[]{35} Sekiguchi M., Fukugita M., 2000, \aj, 120, 1072 
(astro-ph/9904299) 

\bibitem[]{36} van 't Veer-Menneret C., M\'egessier, 1996, \aap, 309, 879

\bibitem[]{40} Westera P., Lejeune T., Buser R., 1999, ASP Conf. Series, Hubeny I., Heap S.R.
and Cornett R.H (eds.), Vol. 192, p.203, (astro-ph/9906064)

\bibitem[]{41} Zakhozhaj V.A., Shaparenko E.F., 1996, Kinematika Fiz.,
Nebesn. Tel., 12, part no 2, 20-29, [ZS96]

\bibitem[]{42} Zorec J., 1992, {\it Hipparcos, Goutelas 1992}, Benest D., Froeschl\'e C. eds., p.407      

\end{thebibliography}
\end{document}